\documentclass[12pt]{iopart}

\usepackage{graphicx}
\usepackage{aas_macros}
\usepackage{hyperref}
\usepackage{bm}
\usepackage{color}
\usepackage{amssymb}
\usepackage{lineno}

\usepackage{caption}
\usepackage{subcaption}

\begin{document}

\title[Towards robust GW detections from individual SMBHBs]{Towards robust gravitational wave detections\\from individual supermassive black hole binaries}

\author{Bence B\'ecsy}
\address{%
Department of Physics, Oregon State University, Corvallis, OR 97331, USA
}
\address{%
Institute for Gravitational Wave Astronomy and School of Physics and Astronomy, University of Birmingham, Edgbaston, Birmingham B15 2TT, UK
}
\ead{b.bence@bham.ac.uk}

\author{Neil J.~Cornish}
\address{%
 eXtreme Gravity Institute, Department of Physics, Montana State University, Bozeman, MT 59717, USA
}%

\author{Polina Petrov}
\address{Department of Physics and Astronomy, Vanderbilt University, 2301 Vanderbilt Place, Nashville, TN 37235, USA}

\author{Xavier Siemens}
\address{Department of Physics, Oregon State University, Corvallis, OR 97331, USA}

\author{Stephen R. Taylor}
\address{Department of Physics and Astronomy, Vanderbilt University, 2301 Vanderbilt Place, Nashville, TN 37235, USA}

\author{Sarah J. Vigeland}
\address{Center for Gravitation, Cosmology and Astrophysics, Department of Physics, University of Wisconsin-Milwaukee,\\ P.O. Box 413, Milwaukee, WI 53201, USA}

\author{Caitlin A. Witt}
\address{Center for Interdisciplinary Exploration and Research in Astrophysics (CIERA), Northwestern University, Evanston, IL 60208}
\address{Adler Planetarium, 1300 S. DuSable Lake Shore Dr., Chicago, IL 60605, USA}


\date{\today}

\begin{abstract}
The recent discovery of the stochastic gravitational-wave background via pulsar timing arrays will likely be followed by the detection of individual black hole binaries that stand out above the background. However, to confidently claim the detection of an individual binary, we need not only more and better data, but also more sophisticated analysis techniques. In this paper, we develop two new approaches that can help us more robustly ascertain if a candidate found by a search algorithm is indeed an individual supermassive black hole binary. One of these is a coherence test that directly compares the full signal model to an incoherent version of that. The other is a model scrambling approach that builds null distributions of our detection statistic and compares that with the measured value to quantify our confidence in signal coherence. Both of these rely on finding the coherence between pulsars characteristic to gravitational waves from a binary system. We test these methods on simple simulated datasets and find that they work well in correctly identifying both true gravitational waves and false positives. However, as expected for such a flexible and simple signal model, confidently identifying signal coherence is significantly harder than simply finding a candidate in most scenarios.
Our analyses also indicate that the confidence with which we can identify a true signal depends not only on the signal-to-noise ratio, but also on the number of contributing pulsars and the amount of frequency evolution shown by the signal.
\end{abstract}



\section{Introduction}
\label{sec:intro}

Gravitational waves (GWs) with nHz frequencies can be best probed by pulsar timing arrays (PTAs), which monitor millisecond pulsars over decades to find the minute imprints of GWs on the time-of-arrivals (TOAs) of the observed radio pulses emitted by these pulsars (for a review see Ref.~\cite{Steve_book}). Analyses of the most recent PTA datasets \cite{nanograv_15yr_data, epta_dr2_data, ppta_dr3_data, MPTA_4p5yr_data} found evidence for the presence of a stochastic GW background (GWB) \cite{nanograv_15yr_gwb, epta_dr2_gwb, ppta_dr3_gwb, cpta_dr1_gwb, MPTA_4p5yr_gwb}. Although we cannot rule out more exotic explanations at this point \cite{nanograv_15yr_new_physics, epta_dr2_astro}, the most likely source of this GWB is a population of inspiralling supermassive black hole binaries (SMBHBs) \cite{nanograv_15yr_astro, epta_dr2_astro}. The loudest of these SMBHs are expected to be individually resolvable not long after observing the GWB \cite{SVV2009, RosadoExpectedProperties, ChiaraLandscape,Luke_single_source,RealisticDetection1}.

Various Bayesian analysis methods have been developed to find and characterize these individual SMBHB signals \cite{NeilCWMethods, Lee_et_al_CW_methods, JustinCWMethods, Steve_accelerated_CW, 2016ApJ...817...70T, QuickCW, FurgeHullam}. Over the years, these methods have used more and more sophisticated signal and noise models and continually improved computational efficiency to keep up with growing PTA datasets. However, further analyses are required to confidently claim a detection of GWs from and individual SMBHB. The primary reason for this is that these methods can only tell us the statistical preference between a model that includes an SMBHB and pulsar noise vs.~a model including only pulsar noise. However, they cannot inform us about how likely it is that the SMBHB model is indeed the correct one. Although a candidate can be confirmed by finding an electromagnetic counterpart, the challenges of that scenario (e.g.~identifying the host galaxy \cite{PolinaHostIdentification2024} and uncertainties in the expected electromagnetic signatures \cite{EM_counterpart_review}) warrant an approach that relies only on GW data. In this paper we develop two such new analysis techniques, which can help us more robustly ascertain if a candidate found by a search algorithm is indeed an SMBHB.

The primary concern when trying to ascertain if a candidate SMBHB is real is that other single-frequency features can masquerade as SMBHBs. Even if just one or a few pulsars' data have a sinusoidal feature, the SMBHB+noise model can be significantly preferred over the noise-only model. What truly distinguishes an SMBHB GW signal from all other sinusoidal features is the specific way the signal appears coherently in each pulsar according to the pulsars' response function. To characterize how confident we are in the presence of this specific pattern, we develop a coherence test (analogous to similar tests used for ground-based GW detector data \cite{Veitch_Vecchio_coherence2010}). This determines the statistical preference between an SMBHB model and a model of incoherent sinusoidal signals in each pulsar. We note that while there are several processes that can give rise to a sinusoidal signal (e.g.~ultralight dark matter \cite{ULDM_Khmelnitsky2014}, asteroids orbiting a pulsar \cite{RossAsteroidBelt}, or solar system ephemeris errors \cite{BayesEphem}) and could be compared to the SMBHB model one-by-one, our approach instead remains agnostic by comparing the SMBHB model to a generic incoherent model. 

Another approach to further scrutinize the SMBHB interpretation of the candidate is to analyze modified versions of the dataset where the data or model is intentionally corrupted in a way that erases the coherence of the signal between pulsars. This is analogous to techniques developed for the GWB \cite{neil_robust_detection,all_correlations_must_die,JeffBackgroundDistribution,DiMarco_scrambling2023,DiMarcoScrambling2024}. Comparing the results from these scrambled datasets with the unscrambled result can tell us how significant the specific correlation structure of the SMBHB signal is. Thus it provides an additional line of evidence for or against the SMBHB interpretation.

The paper is organized as follows. In Section \ref{sec:methods} we introduce our signal model (Section \ref{ssec:signal_model}) and describe how the coherence test (Section \ref{ssec:coherence_test}) and scrambling (Section \ref{ssec:null_distribution}) work. In Section \ref{sec:tests} we test these methods on simulated datasets with either an SMBHB signal or other sinusoidal models that could mimic an SMBHB signal. In Section \ref{sec:conclusion}, we make concluding remarks and describe future directions.

\section{Methods}
\label{sec:methods}

\subsection{Signal model, priors, and sampling}
\label{ssec:signal_model}

A key ingredient in understanding binary origin tests is the signal model that describes GWs emitted by an SMBHB. Here we give a brief overview of the signal model in a formulation that will help us introduce the coherence test in Section \ref{ssec:coherence_test} (for more details see e.g.~Ref.~\cite{FurgeHullam}).
The GW signal of a circular SMBHB in the data of the $\alpha$th pulsar can be written as a sum over four filter functions:
\begin{equation}
    s_{\alpha}^{\mathrm{CW}} = \sum_{i=1}^4 b_{i\alpha} S^i_{\alpha},
    \label{eq:signal_decomp}
\end{equation}
where
\begin{equation}
    S^i_{\alpha}=[\sin (2 \pi f_{\rm E} t); \cos (2 \pi f_{\rm E} t); \sin (2 \pi f_{\alpha} t); \cos (2 \pi f_{\alpha} t)]
\end{equation}
are the four necessary filter functions at the Earth-term ($f_{\rm E}$) and pulsar-term ($f_{\alpha}$) frequencies. Here we make the realistic assumption that there is no frequency evolution during the $\mathcal{O}(10 \ {\rm year})$ observing window, but there can be significant frequency evolution during the $\mathcal{O}(10^3 \ {\rm year})$ light-travel time between Earth and the pulsar. This is a good assumption for most realistic scenarios (see Ref.~\cite{FurgeHullam} for more details). The $b_{i\alpha}$ coefficients\footnote{For an explicit expression, see Ref.~\cite{FurgeHullam}} are responsible for making the signal coherent across all pulsars, as they depend on parameters common to all pulsars, such as the signal amplitude $A_{\rm E}={\cal M}^{5/3} d_L^{-1} (2 \pi f_{\rm E})^{2/3}$, which is determined by the observer-frame chirp mass of the binary ($\mathcal{M}$), the luminosity distance to the source ($d_L$) and the observer-frame Earth term GW frequency ($f_{\rm E}$). The $b^i_{\alpha}$ coefficients are also influenced by the initial phase of the Earth-term signal ($\Phi_0$), the inclination angle of the binary's orbit ($\iota$), the GW polarization angle ($\Psi$), and the $F^{+/\times}_{\alpha}$ antenna patterns, which in turn depend on the sky location of the binary ($\theta$ and $\phi$). In addition, the pulsar terms are influenced by the GW phase ($\Phi_{\alpha}$) and observer-frame GW frequency ($f_{\alpha}$) at the pulsar's location\footnote{In principle, the precise value of $f_{\alpha}$ also determines $\Phi_{\alpha}$. However, it is common practice to parametrize these separately, which helps to avoid hard-to-sample comb-like posteriors that would emerge in the common scenario where pulsar distances are not known within a GW wavelength (see e.g.~Ref.~\cite{JustinCWMethods}).}. Together, these result in a total of $8+2N_{\rm PSR}$ parameters that describe the signal, where $N_{\rm PSR}$ is the number of pulsars in the array.

We use uniform noninformative priors on most of these parameters with the following prior ranges: $\log_{10} A_{\rm E} \in [-18,-11]$, $\log_{10} (f_{\rm E}/1 \ {\rm Hz}) \in [-8.5,-7.5]$, $\log_{10} (\mathcal{M}/1 \ {\rm M_{\odot}}) \in [7,10]$, $\cos \theta \in [-1,1]$, $\cos \iota \in [-1,1]$, $\phi \in [0,2\pi]$, $\Phi_0 \in [0,2\pi]$, $\psi \in [0,\pi]$, $\Phi_\alpha \in [0,2\pi]$. Instead of the pulsar term frequency, we sample the pulsar distance parameter, and we set a Gaussian prior on that centered on the true value. We use a hyper-fast interpolated likelihood that is marginalized over pulsar phase and distance parameters \cite{FurgeHullam}. This is implemented in the \texttt{FurgeHullam}\footnote{Publicly available at: \url{https://github.com/bencebecsy/FurgeHullam}} package, and after a short setup it achieves millisecond-scale evaluation times. In addition, since it provides the likelihood already marginalized over $2N_{\rm PSR}$ parameters, we can use nested sampling (see e.g.~Ref.~\cite{nested_sampling}) over the remaining 8 parameters with the \texttt{dynesty} package \cite{dynesty}. This results in robust evidence calculations in $\mathcal{O}(5 \ {\rm min})$.

\subsection{Signal coherence test}
\label{ssec:coherence_test}

There are usually two models considered in an SMBHB GW search: the noise-only model (NOISE), and a model that in addition to the noise model, also includes the continuous-wave GW signal from an SMBHB (CW). The idea of the coherence test is to introduce an incoherent version of the signal model (INCOH). Such a model will be a generalized version of the CW model, with more flexibility allowing it to fit a wider range of "signals". As such, it will always achieve at least as good a fit as the CW model, and if the data indeed only contains an individual SMBHB, the two models will fit equally well. However, this increased flexibility is automatically penalized in the Bayesian analysis, which not only considers goodness-of-fit, but also has a preference towards simpler, more predictive models. This ensures that if both CW and INCOH models fit equally well, CW is preferred in a Bayesian analysis.

To quantify the preference between the CW and INCOH models we can calculate the Bayes factor between these ($\mathrm{BF_{CW-INCOH}}$). Note that since Bayes factors are ratios of evidences, $\mathrm{BF_{CW-NOISE}}=\mathrm{BF_{CW-INCOH}} \mathrm{BF_{INCOH-NOISE}}$. Thus we can think of this as breaking down how much of the total support for the signal model relies on the coherence of the signal. This idea is inspired by a similar approach in LIGO-Virgo data analysis, where the coherent binary black hole model is compared with a model in which the binary can have different parameters in each detector \cite{Veitch_Vecchio_coherence2010}. This is also analogous with the approach used in the PTA GWB analysis, where instead of comparing to the noise-only model, the GWB model is usually compared to a common uncorrelated red noise (CURN) model, which is only different from the GWB model in that it does not have inter-pulsar correlations (see e.g.~Ref.~\cite{neil_robust_detection}).

To construct such an incoherent model, we replace the expression for the signal in Eq.~(\ref{eq:signal_decomp}) with a simple sine-wave:
\begin{equation}
    s_{\alpha}^{\mathrm{INCOH}} = A_{\alpha} \sin (2\pi f_0 t - \Phi_{\alpha}^{'}),
    \label{eq:incoh_waveform}
\end{equation}
where the $A_{\alpha}$ amplitudes and $\Phi_{\alpha}^{'}$ phases are chosen independently for each pulsar, while the $f_0$ frequency is common to all pulsars. Note that when the SMBHB is slowly evolving and thus $f_{\rm E} \simeq f_{\alpha}$ for all pulsars, the incoherent model in Eq.~(\ref{eq:incoh_waveform}) is able to produce the same waveform as the coherent model in Eq.~(\ref{eq:signal_decomp}). However, due to the larger flexibility of the INCOH model, we expect the Bayesian analysis to prefer the more predictive CW model over the INCOH model, when the data includes a true SMBHB signal. This is due to the built-in Occam penalty in Bayesian analysis that prefers simpler models.

It is important to choose priors consistently between the coherent and incoherent models. This simply means choosing the same log uniform prior on $f_0$ as we do on $f_{\rm E}$\footnote{Note that consistency in priors also motivates our choice of requiring the INCOH model to have a common frequency in all pulsars, since allowing the frequency to be an independent free parameter in all pulsars would result in the INCOH model having a significantly larger prior volume compared to the CW model.}, and simply using a uniform prior between 0 and $2\pi$ on $\Phi_{\alpha}^{'}$. Setting the prior on $A_{\alpha}$ is more complicated. Although the coherent model has a log-uniform prior on the $A_{\rm E}$ amplitude, we cannot simply adopt that same prior for $A_{\alpha}$. The problem is that having uninformative log-uniform priors on dozens of amplitude parameters corresponding to the dozens (if not hundreds) of pulsars in current PTAs makes this model too flexible. To fit a common signal, the amplitude for each pulsar would need to be set completely independently, which would result in this model being practically always disfavored over the coherent model\footnote{This is the same reason why the frequency is chosen to be the same among all pulsars.}. Instead we are aiming to describe a model where each pulsar have a signal with an amplitude of the same order of magnitude, but with the freedom to set the exact amplitude in each pulsar independently. To mathematically describe this we use a half-normal prior on the amplitude with zero mean and a variance set by the hyperparameter $\mathcal{A}$, i.e.~$A_{\alpha} = | x |$, with $x \sim N(0,\mathcal{A}^2/(2\pi f_0)^2)$, where the frequency factor converts strain amplitude to residuals. We set a uniform prior on the hyperparameter with the same range as for the strain amplitude in the coherent model, i.e.~$\mathcal{A} \in [-18,-11]$. This means that the prior volumes of the coherent and incoherent models have the same dependence on the amplitude prior range, thus if we use the same range, the resulting Bayes factor is independent of the particular prior range used. Note that ideally one would want to use the same prior on $A_{\alpha}$ as the implicitly implied prior on per-pulsar amplitudes in the coherent model. We explore these in \ref{sec:implied_prior}, but for simplicity we use the half-normal distribution as a fast and easy-to-implement prior in this paper. The effects of prior choice will be explored more in future work.

\subsection{Creating null distributions}
\label{ssec:null_distribution}
Another way to quantify our preference for a truly coherent signal is by analyzing the data while either the model or the data are scrambled in a way that is expected to erase correlations between pulsars. By repeating this process many times, we can produce a null distribution of our detection statistic (e.g.~$\mathrm{BF_{CW-NOISE}}$). Comparing the unscrambled (foreground) detection statistic to this null distribution results in a p-value that quantifies our confidence in the signal being coherent.

This idea is used extensively for LIGO-Virgo transient signals, where the coherence can be completely broken by shifting the detectors data by more than the maximum light-travel-time between detectors (see e.g.~Ref.~\cite{ligo_timeslides}). Methods have also been developed to quantify the significance of a GWB signal in PTA data \cite{neil_robust_detection, all_correlations_must_die} by either sky scrambling or phase shifting. Sky scrambling replaces the true sky location associated with each pulsar in the model with a location randomly drawn from a uniform distribution on the sky. Phase shifting adds random phases to the signal model in each pulsar. As a result, both of these approaches kill the cross-pulsar correlations expected from a true GWB signal.

In this work, we apply analogous sky scrambling and phase shifting approaches to the problem of an individual SMBHB detection. Phase shifting works by simply shifting the entire signal (both Earth and pulsar terms) by a phase that we draw independently for each pulsar. Note that while the pulsar terms have their own phase parameters ($\Phi_{\alpha}$), the Earth term only has one global phase parameter ($\Phi_0$). This means that we can only expect phase shifting to affect half of the signal, since the pulsar terms can always correct themselves for a random phase shift.

Sky scrambling also works similarly to the GWB methods, with the added complexity that pulsar sky locations actually appear at two different places in the SMBHB signal model. They appear in the $F^{+/\times}_{\alpha}$ antenna patterns that determine how strongly the two GW polarizations appear in each pulsar. They also affect the time delays between the Earth and pulsar terms which determine the difference between the $f_{\rm E}$ and $f_{\alpha}$ frequencies. We found that unless the binary evolves unrealistically fast, it does not matter if the time delay is affected by the scramble or not. This is because for most systems that we expect to detect, the evolution is detected much less significantly compared to the overall signal. Interestingly, for high-SNR fast-evolving systems, scrambling only in the antenna patterns results in slightly higher significances, because such a scramble results in inconsistent sky locations in different parts of the model, which makes it even harder to find a good fit for the data. For the rest of the paper, we always scramble sky locations everywhere they appear in the model.

In addition to this well-established sky scrambling method, we also introduce a new variant we call \emph{sky shuffling}. The idea is that instead of randomly drawing locations uniformly in the sky, we randomly exchange (shuffle) the locations assigned to each pulsar. Unlike sky scrambling which makes any PTA isotropic, sky shuffling retains the anisotropic configuration of the given PTA. As we will see in Section \ref{ssec:scrambling_results}, this can be important in some scenarios. 

\section{Tests on simulated datasets}
\label{sec:tests}

We tested the methods described in Section \ref{sec:methods} on simple simulated datasets. These consist of 10 or 15 pulsars uniformly distributed on the sky. All pulsars' data are uniformly sampled every 30 days for a total timespan of 10 years and have the same white noise level of 0.5 $\mu$s. We do not include any red noise or GWB signal. Our methods do not make use of these simplifications and can trivially be applied to realistic or even real PTA data. However, the simplicity of the datasets meant that we could explore many different scenarios and the results are easier to interpret.

\subsection{Coherence test}
\label{ssec:coherence_results}

\begin{figure}[htbp]
\includegraphics[width=\columnwidth]{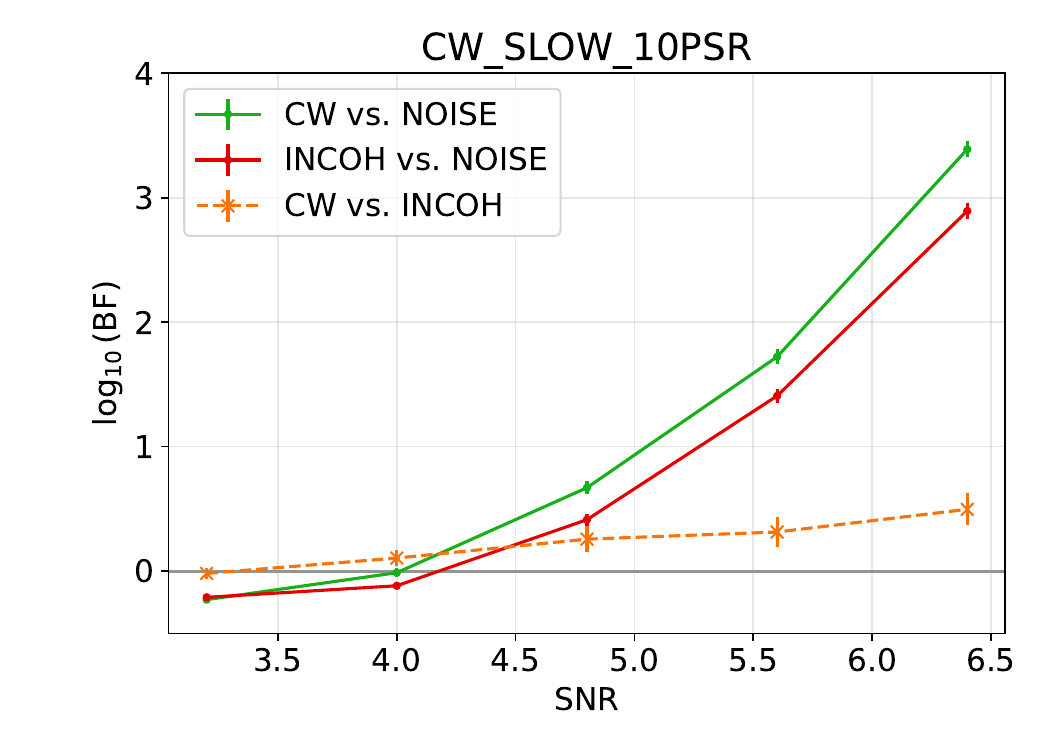}
\caption{Coherence test on 10 PSR dataset with slowly evolving SMBHB signal. Bayes factors between the three models of interest are shown as a function of the SNR. Note that while the preference for both coherent and incoherent model against the noise-only model increases with SNR, the Bayes factor between coherent and incoherent model also increases. This indicates that for stronger signals we can more confidently identify the correct model.}
\label{fig:bfs_10psr_slowly_evolving}
\end{figure}

Figure \ref{fig:bfs_10psr_slowly_evolving} shows the results of the coherence test on 10-pulsar simulated datasets with a slowly evolving CW signal of various signal-to-noise ratios, SNRs (CW\_SLOW\_10PSR). Other parameters of this signal are shown in Table \ref{tab:simulations}, where a fiducial amplitude is shown which we varied to achieve a range of SNRs. We can see in Figure \ref{fig:bfs_10psr_slowly_evolving} that both $\mathrm{BF_{CW-NOISE}}$ and $\mathrm{BF_{INCOH-NOISE}}$ increase with the SNR of the simulated signal. This highlights that even a model that is not true (in this case \emph{INCOH}) can be strongly favored over the noise-only model. However, as expected, the correct \emph{CW} model is favored over the \emph{INCOH} model, and this preference gets stronger with increasing signal amplitude, as indicated by the increasing trend in $\mathrm{BF_{CW-INCOH}}$.

\begin{table*}[htbp]
\centering
\caption{%
Binary parameters in simulated datasets. Amplitudes listed are the ones used in Section \ref{ssec:scrambling_results}. Results in Section \ref{ssec:coherence_results} used a range of amplitudes around this value to explore a wide variety of Bayes factors. \label{tab:simulations}
}
\begin{tabular}{|l|cccccccc|}
\hline
 & $A_{\rm E}$ & $f_{\rm E}$ [nHz] & ${\cal M}$ [$M_{\odot}$] & $\theta$ & $\phi$ & $\iota$ & $\Phi_0$ & $\Psi$ \\
\hline
CW\_SLOW\_10PSR & $3.5\times10^{-15}$ & 8.0 & $10^7$ & 1.4 & 3.3 & 0.7 & 2.0 & 0.3\\
\hline
CW\_SLOW\_15PSR & $2.5\times10^{-15}$ & 8.0 & $10^7$ & 1.4 & 3.3 & 0.7 & 2.0 & 0.3\\
\hline
CW\_FAST\_10PSR & $1.4\times10^{-14}$ & 20.0 & $6 \times 10^9$ & 1.4 & 3.3 & 0.7 & 2.0 & 0.3\\
\hline
\end{tabular}
\end{table*}

\begin{table*}[htbp]
\centering
\caption{%
Monopole signal parameters in simulated datasets. Amplitudes listed are the ones used in Section \ref{ssec:scrambling_results}. Results in Section \ref{ssec:coherence_results} used a range of amplitudes around this value to explore a wide variety of Bayes factors. \label{tab:simulations_mono}
}
\begin{tabular}{|l|ccc|}
\hline
 & $A_\alpha$ [s] & $f_0$ [nHz] & $\Phi'_\alpha$ \\
\hline
MONO\_10PSR & $1.8\times10^{-7}$ & 8.0 & 3.9\\
\hline
MONO\_15PSR & $1.4\times10^{-7}$ & 8.0 & 3.9\\
\hline
\end{tabular}
\end{table*}

The example in Figure \ref{fig:bfs_10psr_slowly_evolving} suggests that it is significantly harder to ascertain the coherence of a signal than it is to simply find a candidate preferred over the noise-only model. This is akin to the GWB detection problem, where the Bayes factor between a common signal and noise-only is usually many orders of magnitude larger than the Bayes factor between a correlated and an uncorrelated signal\footnote{For example, the analysis of the NANOGrav 15-year data finds a Bayes factor of $\sim 10^{12}$ between a common signal and noise-only, but only a Bayes factor of 200-1000 in favor of correlations \cite{nanograv_15yr_gwb}.}. However, we expect this to be dependent on properties of the signal and the PTA we observe it with. For example, coherence will be easier to ascertain if a larger number of pulsars contribute to the detection. We can also expect the complexity of the signal (fast vs.~slow evolving) to play an important role. Fast evolution results in a unique feature where each pulsar shares a common frequency corresponding to the Earth term, but all of them have a lower-frequency component as well at various different frequencies corresponding to the pulsar terms. This increased complexity will make it easier to distinguish between CW and INCOH models. To explore these effects, we analyzed two additional dataset collections. One with the same signal parameters as before, but with 15 pulsars instead of 10 (CW\_SLOW\_15PSR), and another one with higher chirp mass and frequency so that there is a clear frequency difference between the Earth and pulsar terms (CW\_FAST\_10PSR). The list of parameters for these datasets is described in Table \ref{tab:simulations}, except for the amplitude which was tuned to get datasets with roughly the same range of $-0.5 \lesssim \log_{10} \mathrm{BF_{CW-NOISE}} \lesssim 4.5$.

\begin{figure}[htbp]
\includegraphics[width=\columnwidth]{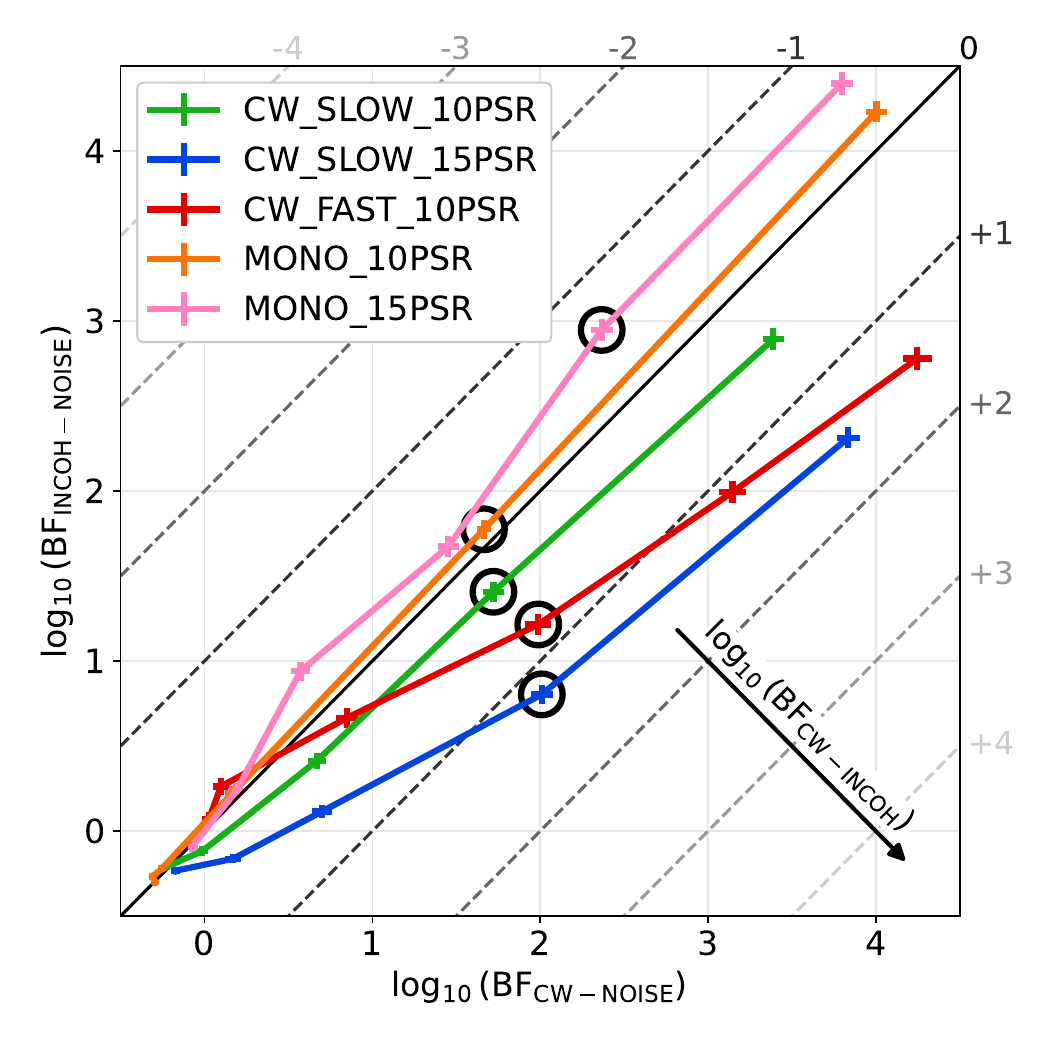}
\caption{Bayes factors between the \emph{CW} and \emph{NOISE} models vs Bayes factors between the \emph{INCOH} and \emph{NOISE} models for various datasets (see details in main text and Tables \ref{tab:simulations} and \ref{tab:simulations_mono}). Bayes factors between the \emph{CW} and \emph{INCOH} models increase from top left to bottom right as indicated by the arrow, and and constant values are shown by dashed lines. We can see that datasets with an injected CW are all below the diagonal, indicating a correct preference for the coherent \emph{CW} model. Datasets with a monopolar sine wave injection are all above the diagonal indicating a correct preference for the incoherent \emph{INCOH} model. Black circles highlight datasets used for the scrambling tests in Section \ref{ssec:scrambling_results}.}
\label{fig:coherence_test_bf_summary}
\end{figure}

Figure \ref{fig:coherence_test_bf_summary} shows $\mathrm{BF_{CW-NOISE}}$ and $\mathrm{BF_{INCOH-NOISE}}$ for these three collections of datasets: the datasets from Figure \ref{fig:bfs_10psr_slowly_evolving} (green), the 15-pulsar datasets (blue), and the fast evolving datasets (red). We can see that all three of these show that $\mathrm{BF_{CW-NOISE}}$ and $\mathrm{BF_{INCOH-NOISE}}$ grow together, but as we have seen on Figure \ref{fig:bfs_10psr_slowly_evolving}, $\mathrm{BF_{CW-NOISE}}$ grows faster, so we can eventually distinguish the \emph{CW} and \emph{INCOH} models. Note that points below the main diagonal indicate preference for the \emph{CW} model, while points above it prefer the \emph{INCOH} model. Dashed lines indicate contours of constant $\log_{10} \mathrm{BF_{CW-INCOH}}= \{ -4,-3,-2,-1,0,1,2,3,4 \}$. We can see that the 15PSR line (blue) always goes below the 10PSR line (green), which indicates that at a given level of detection confidence (i.e.~$\mathrm{BF_{CW-NOISE}}$), more pulsars help better distinguish the coherent and incoherent models. Also note that the fast evolving line (red) starts out to be above the slow evolving line (green), but at higher $\mathrm{BF_{CW-NOISE}}$ it goes below it. This shows that the coherence of evolving systems is easier to establish as long as they are detected with a high enough confidence that the pulsar terms make a significant contribution. Thus we can conclude that the coherence test is able to correctly identify the right model for SMBHBs, and the confidence of this identification increases with the detection confidence (or SNR), the number of contributing pulsars, and signal complexity (frequency evolution).

Another important question is whether the coherence test can correctly dismiss candidates that are in fact not SMBHBs. It is impossible to examine this question exhaustively, because there are infinite ways a candidate can deviate from the SMBHB signal model. In fact, if the deviations are small enough, we will never be able to distinguish them. Thus we limit ourselves to a simple example here: a purely monopolar sine wave, i.e.~a sinusoidal signal with the exact same amplitude, frequency, and phase in each pulsar. We can consider this a worst case scenario, because such a candidate would actually show some form of coherence, just not the right kind of coherence expected for a CW signal. Also note that this is not the same as the incoherent model used for the coherence test, because that one allows independent phases and amplitudes in each pulsar. This is important, because it assures that we test our approach with a signal that matches neither our CW nor our INCOH model perfectly. We show results for such datasets on Figure \ref{fig:coherence_test_bf_summary} with 10 pulsars (MONO\_10PSR, orange) and 15 pulsars (MONO\_15PSR, pink). Parameters of the injected signals in these datasets are listed in Table \ref{tab:simulations_mono}. We can see that all of these points lie above the main diagonal, indicating that the incoherent model is preferred. Note however, that even at the strongest signals examined, this preference is not strong. This is due to the fact that this monopolar sine wave is inherently hard to distinguish from an SMBHB signal. However, it is still reassuring to see that the coherence test can correctly reject these candidates, even if with only mild confidence.

\subsection{Scrambling}
\label{ssec:scrambling_results}

Compared with the coherence test discussed in Section \ref{ssec:coherence_results}, the scrambling approach is significantly more computationally expensive. This is because here we need to perform the same Bayesian analysis on hundreds of scrambled datasets compared to just one coherent and one incoherent analysis in the case of the coherence test. In fact, the more significant the candidate is, the more runs we need, with at least $N$ runs required to measure a p-value of $1/N$. Due to this, we will only analyze a single dataset from each collection. These are marked by black circles in Figure \ref{fig:coherence_test_bf_summary}, and their parameters are shown in Tables \ref{tab:simulations} and \ref{tab:simulations_mono}.

\begin{figure}[!htbp]
 \centering
            \includegraphics[width=1.0\linewidth]{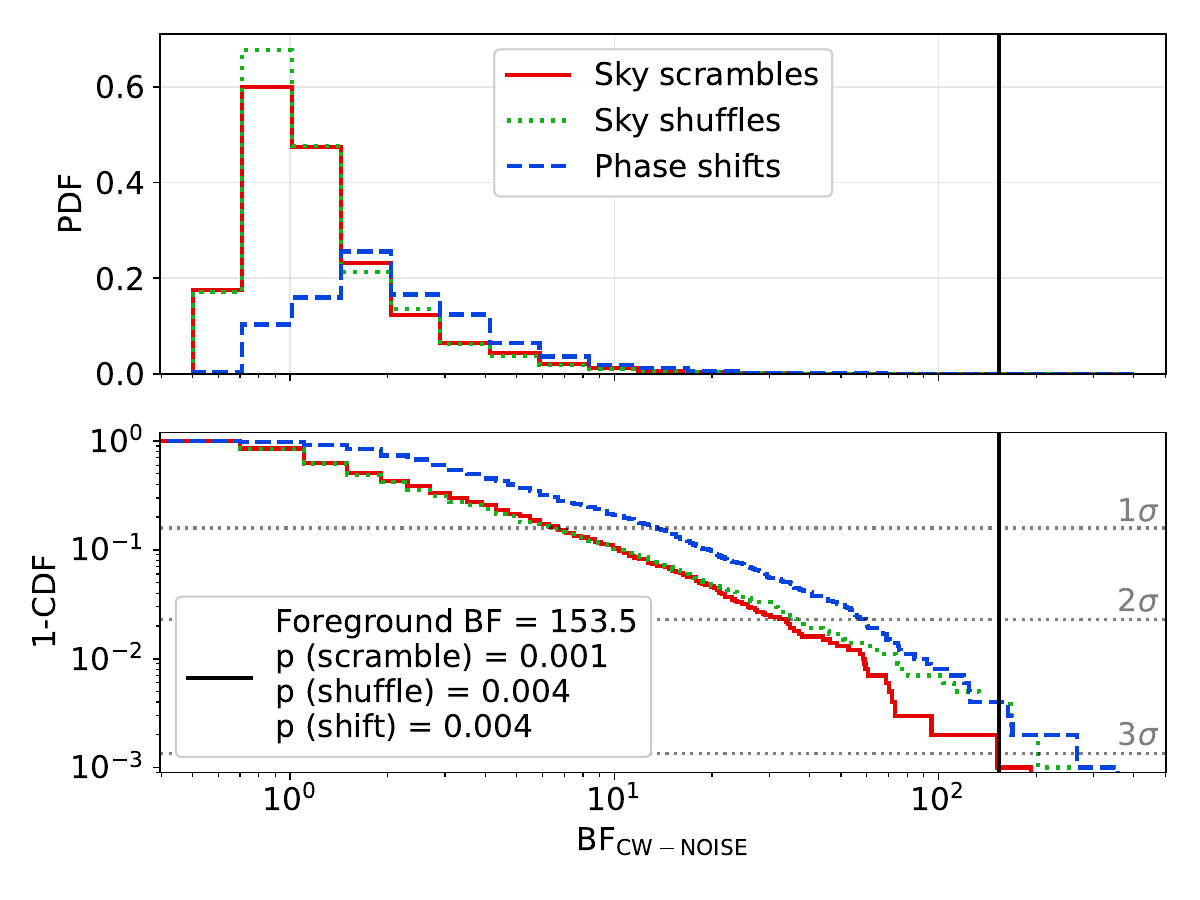}
 \caption{Probability density function (top) and cumulative distribution function (bottom) of Bayes factors found in sky scrambled (red solid lines), sky shuffled (green dotted lines), and phase shifted (blue dashed lines) versions of the CW\_SLOW\_15PSR dataset. We also show the Bayes factor value found in the unscrambled dataset (black vertical lines), which lie well in the tail of the distribution. Horizontal dotted lines indicate p-values corresponding to 1$\sigma$, 2$\sigma$, and 3$\sigma$. The resulting p-values that can be read off the bottom panel are also shown. Note that these can be highly dependent on the method used to create the distribution.}
 \label{fig:background_cw_fast_evolve_15psr}
\end{figure}

Figure \ref{fig:background_cw_fast_evolve_15psr} shows the null distribution we get with the three different scrambling methods for the CW\_SLOW\_15PSR dataset (see Table \ref{tab:simulations}). It also shows the Bayes factor we get from the unscrambled analysis as vertical black lines. The top panel shows the probability density function of the null distribution. We can see that all three methods are effective in erasing correlations, and the BF from the unscrambled analysis is in the tail of these distributions. Phase shifting is noticeably less effective than sky scrambling and sky shuffling. This is due to the fact that while phase shifting only affects the phasing of the signal, sky scrambles and shuffles affect both the phasing and the amplitude. Due to the free pulsar term phases in the model, the effect of scrambling on the signal amplitude is more important than their effect on the phase. The bottom panel of Figure \ref{fig:background_cw_fast_evolve_15psr} shows corresponding extinction curves, which are defined as 1 minus the cumulative distribution function (CDF). This is a useful representation, because the p-values can be easily read off by simply finding the value of 1-CDF at the unscrambled Bayes factor value. We can see that the sky scramble (p-value$\ = 0.001$) indicate higher significance compared to both the sky shuffle (p-value$\ =0.004$) and the phase shift (p-value$\ =0.004$), but all three p-values are around the 3-$\sigma$-equivalent value of 0.0013. We can see that the null distribution and resulting p-values depend on the method of scrambling used. This suggests that these are not rigorously defined p-values associated with the null hypothesis of an uncorrelated signal. However, they are still useful as rough indicators of signal coherence.

\begin{figure}[!htbp]
 \centering
    \begin{subfigure}[b]{1.0\textwidth}
            \centering
            \includegraphics[width=0.75\linewidth]{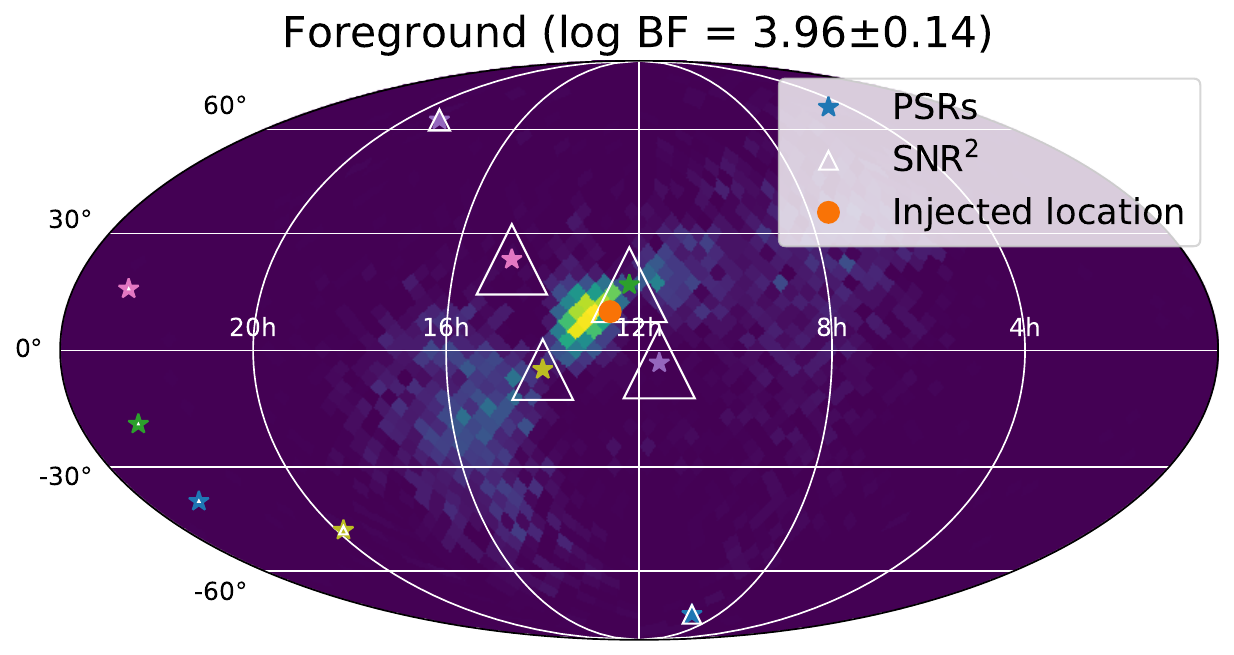}
    \end{subfigure}%
    \hfill
    \begin{subfigure}[b]{1.0\textwidth}
            \centering
            \includegraphics[width=0.75\linewidth]{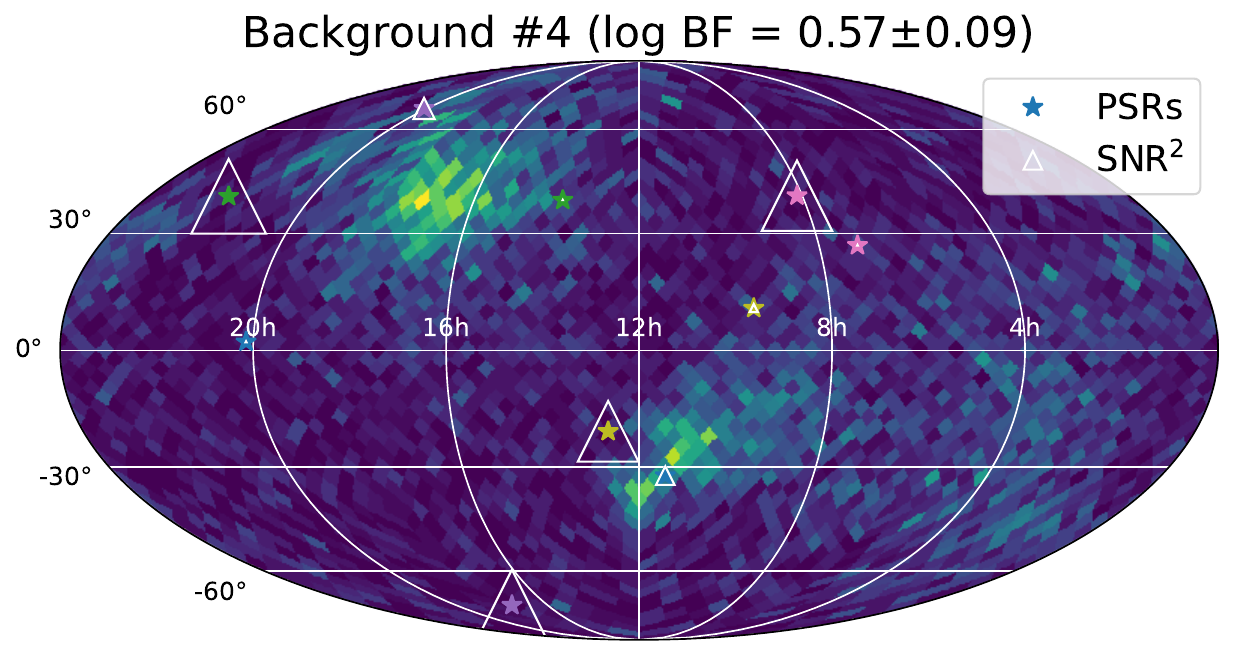}
    \end{subfigure}%
    \hfill
    \begin{subfigure}[b]{1.0\textwidth}
            \centering
            \includegraphics[width=0.75\linewidth]{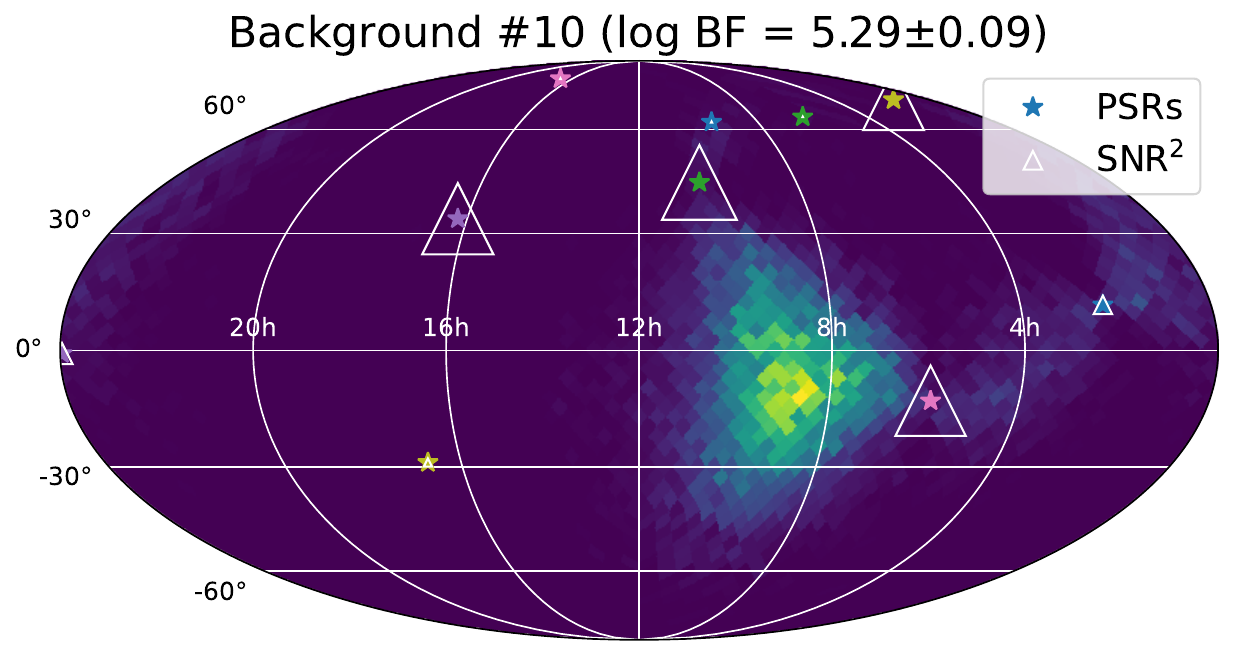}
    \end{subfigure}%
    \hfill
 \caption{Sky location posteriors for unscrambled analysis (top panel), and scrambled analysis with low (middle panel) and high (bottom panel) Bayes factors. Pulsar locations are also shown with consistent color coding, so one can follow where a particular pulsar got moved in a given scramble. The square of the per-pulsar SNR in each pulsar is indicated by the white triangles' sizes. We can see that scrambles can retain a high Bayes factor if by chance high-SNR pulsars are positioned with no low-SNR pulsars between them.}
 \label{fig:skymaps}
\end{figure}

To illustrate how sky scrambling can eliminate a coherent signal, we show sky location recoveries for a few examples on Figure \ref{fig:skymaps}. All of these are for the CW\_SLOW\_10PSR dataset. The top panel corresponds to the unscrambled foreground, the middle panel for a particular scramble realization where the CW significance is greatly diminished, and the bottom panel is for a realization that has a significance even higher than the foreground. The top panel also shows the true injected location of the source (orange dot), which is recovered well. Each panel shows the location of the 10 pulsars with a consistent color coding, so that we can follow where a pulsar from the unscrambled dataset ends up on the sky in different scrambles. To help us understand how strongly the signal appears in each pulsar, white triangles are shown with sizes  proportional to SNR$^2$ in the given pulsar\footnote{Note that this is the SNR the pulsar sees this particular signal with, and not a general indicator of that pulsar's sensitivity, which is uniform across all pulsars in this dataset.}. We can see that the signal is strongest in the four pulsars closest to the source in the unscrambled dataset. This makes sense as the antenna pattern falls off quickly with angular distance from the source. In the middle panel, these pulsars end up far away from each other on the sky, with low-SNR pulsars between them. This means that we cannot find any sky location where the signal can appear in pulsars the same way it does in the unscrambled scenario. This result in an unconstrained sky location recovery, and a low Bayes factor. On the other hand, in the example shown in the bottom panel, the four key pulsars (green, pink, purple, yellow) are all relatively close together with no other pulsars in between. As a result, we are able to assign the signal a sky location where it can appear in these pulsars similar to how it appears in the unscrambled dataset. Thus, even though the recovered sky location is completely different than the true location, the significance can be unaffected, or even higher than the foreground, as we see in this particular example. Based on this example, we can expect the scrambling to work better with a larger number of pulsars, as it gets increasingly rare to get sky locations where the high-SNR pulsars are clustered close together by chance.

\begin{figure}[htbp]
\includegraphics[width=\columnwidth]{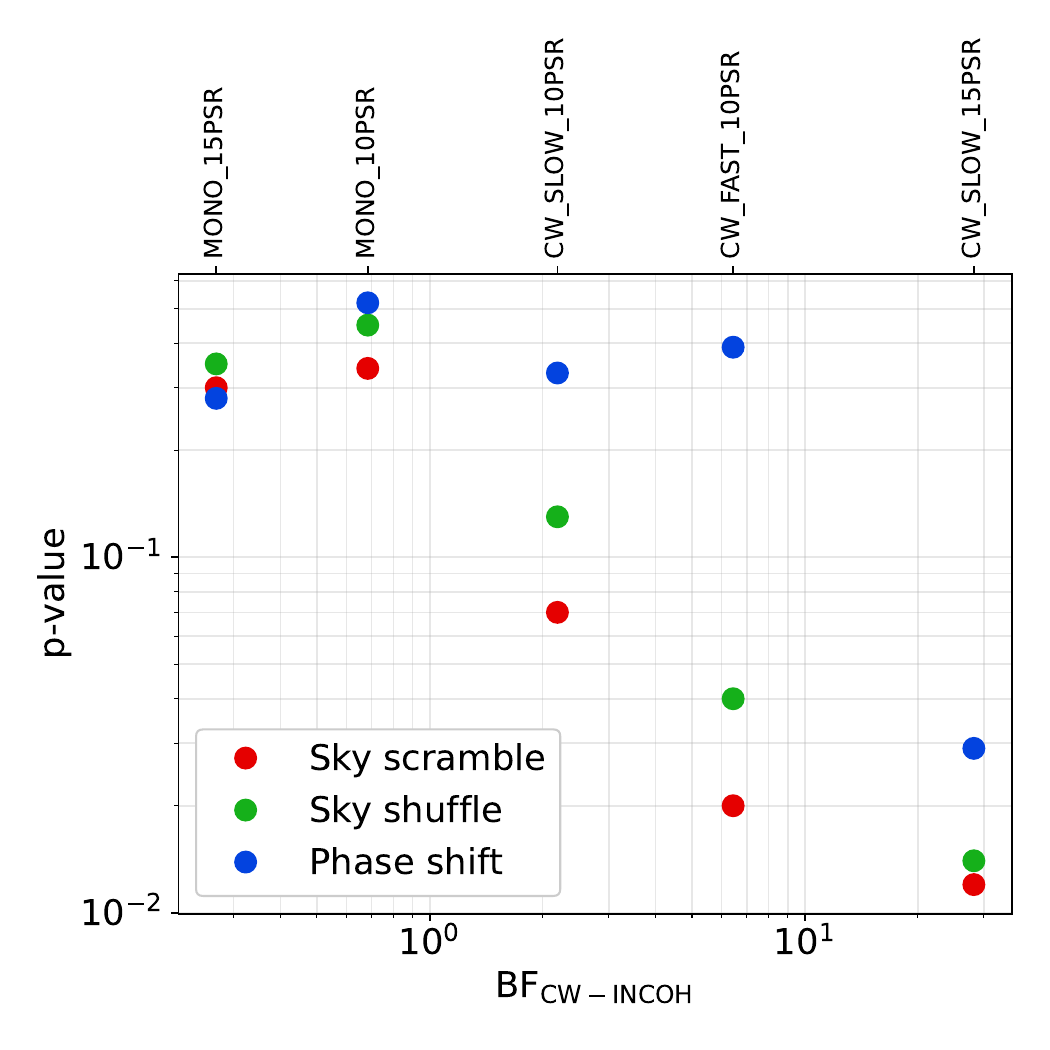}
\caption{Comparison between coherence test and scrambling methods. Shown are the p-values calculated with all three methods for all five datasets in Table \ref{tab:simulations} and their corresponding coherent vs.~incoherent Bayes factors. Note the negative correlation between these, indicating that the two approaches tend to agree in their assessment of signal coherence. Also note that while the Bayes factor can go arbitrarily low to express a strong preference against the CW model, p-values flatten out and give no indication on how strongly the CW model is disfavored.}
\label{fig:bf_vs_pvalue}
\end{figure}

We analyzed all three SMBHB datasets from Table \ref{tab:simulations} and also both monopole datasets from Table \ref{tab:simulations_mono}. The resulting p-values are shown in Figure \ref{fig:bf_vs_pvalue} as a function of $\mathrm{BF_{CW-INCOH}}$. We can see that candidates with higher coherent vs.~incoherent Bayes factors tend to have lower p-values. This is reassuring, since these two different approaches are trying to answer the same question of whether the candidate actually shows signs of a GW-like coherence across pulsars. Also note that most datasets show a hierarchy between scrambling methods, with sky scrambling (phase shifting) usually resulting in the lowest (highest) p-values. This is consistent with the already discussed explanation that sky scrambling does better because it affects not only the phasing, but also the amplitude of the signal. It is also important to note that while the coherence test can give a degree of significance when the coherent model is disfavored, the scrambling methods flatten out around a p-value of 0.3-0.6 and give no information on how strongly the CW model is disfavored. 

\begin{figure}[htbp]
\includegraphics[width=\columnwidth]{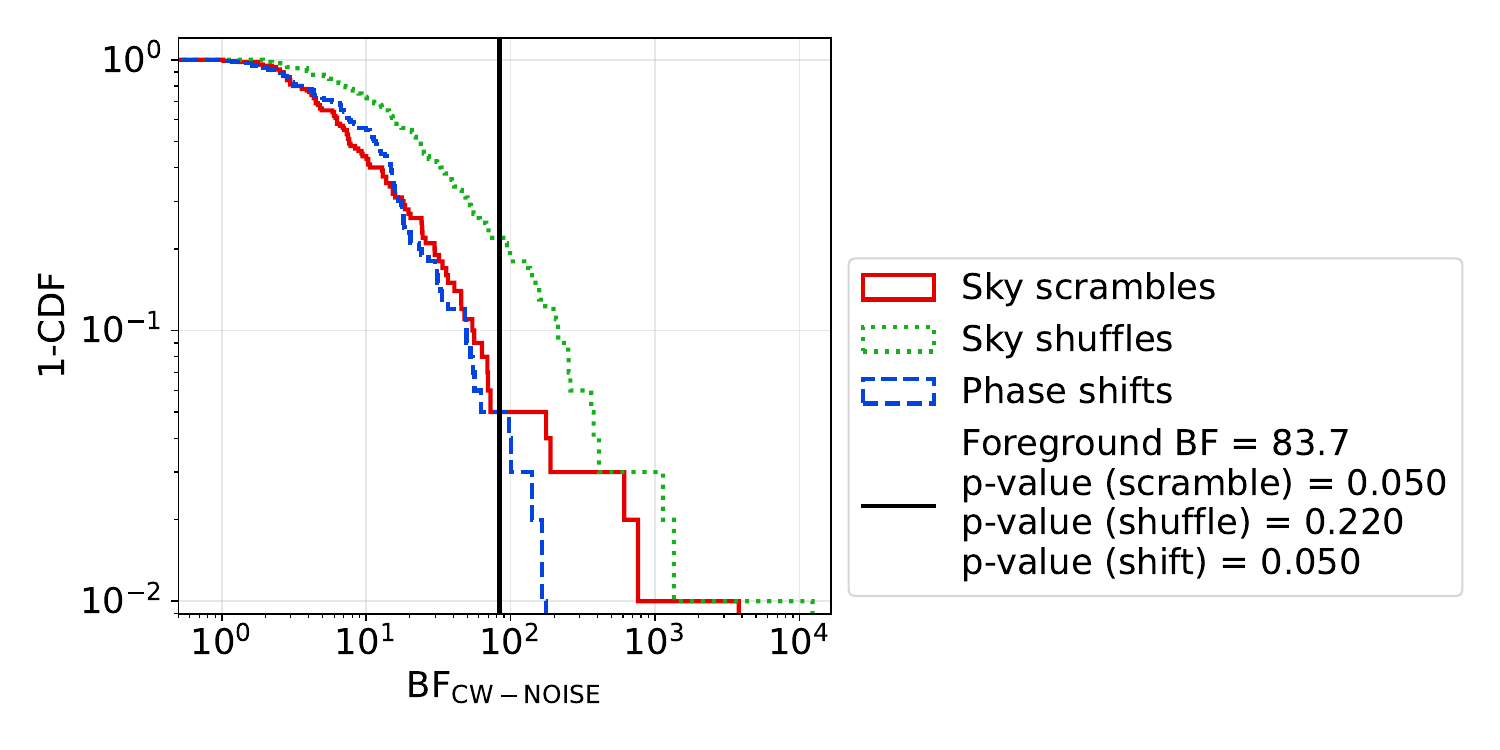}
\caption{Background distributions for monopolar sine-wave injection in realistic NANOGrav 12.5yr-like simulated dataset. The BF found in the unscrambled analysis is shown by the black vertical line. Note that sky scrambles and phase shifts result in erroneously low p-values that indicate a marginally significant SMBHB. This is due to the fact that these destroy the anisotropy of the array. Sky shuffles are able to correctly reject this candidate thanks to their property of preserving the PTA anisotropies.}
\label{fig:bakcground_12p5yr_monopole}
\end{figure}

To highlight the important difference between sky scrambles and sky shuffles, we also analyzed a more realistic dataset based on the NANOGrav 12.5yr dataset \cite{nanograv_12p5yr_data}. We injected white and red noise according to the best-fit values found in the real dataset. We also added a monopolar sine wave signal to the timing residuals with a frequency of 8 nHz and an amplitude of 8 ns, with the same phase in each pulsar. Figure \ref{fig:bakcground_12p5yr_monopole} shows the null distributions we find for this dataset, along with the foreground BF and p-values. We can see that sky scrambles and phase shifts find a relatively low p-value, which might result in us incorrectly claiming a significant SMBHB detection. This is due to the fact that these scrambling methods destroy the anisotropic distribution of pulsars in our PTA. In the unscrambled version, the model is able to put a binary far away from most pulsars on the sky, which results in a roughly monopolar signal. However, once the sky locations are scrambled and the pulsars' distribution on the sky is uniform, no such sky location can be found for the source. The result is systematically lower Bayes factor values in scrambled datasets, resulting in an erroneously low p-value. Sky shuffles on the other hand just exchange sky locations between pulsars, thus preserving their anisotropic distribution. As a result, they are able to correctly reject this candidate as no significant correlations are found.

\section{Conclusion and future work}
\label{sec:conclusion}

Detecting a supermassive black hole binary with pulsar timing arrays will not only confirm the origin of the stochastic gravitational wave background, but also enable many future studies including tests of general relativity, probing supermassive black hole formation and growth, and more. However, the confident detection of such a binary will require additional and better data; as well as more sophisticated detection methods and validation techniques. The latter is particularly important because many of the noise sources in pulsar timing array data are not well understood, and it is challenging to distinguish the simple signal model for a black hole binary from noise features. In this paper, we take the first steps towards this goal by investigating analysis approaches that can inform such a validation process by increasing or decreasing our confidence in a binary candidate solely based on the pulsar timing array data available.

The simplest approach to detection is to carry out a model comparison between a noise-only model and a model with both noise and a gravitational wave signal included. If the model including a signal is preferred, one might claim a detection of that signal. The problem with this simplistic approach is that it only tells us about those two models in question, but nothing about the infinitely many other potential models. As such, it is possible that there are viable models not including a gravitational wave signal that match our data even better. In practice, we cannot test all possible alternative models, so we need creative ways to put the gravitational wave model under further scrutiny. In this paper we propose and test two such approaches.

The first approach is a coherence test, which works by introducing a third model, one similar to the gravitational wave model except that it does not require coherence between different pulsars. If such a model is favored over the gravitational wave model, we should be wary of the gravitational wave interpretation. On the other hand, if the gravitational wave model is preferred, we have more confidence that the data prefer the specific inter-pulsar correlations, and not just the general shape or frequency content of the signal model. This is motivated by similar tests used for the stochastic background in pulsar timing arrays \cite{neil_robust_detection} and for transient signals in ground-based gravitational-wave detectors \cite{Veitch_Vecchio_coherence2010}. We tested this approach on simple simulated datasets and found that it is able to give support for true binary signals and correctly disfavor some examples of false positives. We also found that the confidence with which we can identify a true signal depends not only on the signal-to-noise ratio but also hinges strongly on the number of pulsars contributing to the detection and the amount of frequency evolution shown by the signal. This, in turn, can also inform the strategy for future observing campaigns, as we may benefit from a large number of pulsars as opposed to intensive observations of a few very well-timed ones, if we want to confidently identify a candidate as a true binary.

The second approach is to build a null distribution for our detection statistic (the Bayes factor between noise+binary and noise-only models) under the hypothesis that there are no correlations between pulsars. The result from the main analysis can be compared with this distribution to quantify our confidence in the presence of those correlations that are expected to be there for a gravitational wave signal. This is motivated by similar methods used for the stochastic background \cite{neil_robust_detection, all_correlations_must_die}, and we use similar techniques to build such null distributions, namely phase shifting and sky scrambling. In addition, we also introduce sky shuffling, which exchanges pulsar sky locations instead of randomly drawing new locations uniformly on the sky. We test these on simple simulated datasets and find that in general they work well in correctly identifying both true gravitational waves and false positives. We find that phase shifting only works when there is a sufficiently large number of contributing pulsars. We also find that sky scrambling and phase shifting can fail to reject a monopolar sine-wave false positive in realistic anisotropic pulsar timing arrays. This is due to the fact that these destroy the anisotropy in these arrays, and the sky shuffling method we introduced solves this problem.

We find that the null distribution approach and the coherence test tend to agree in their results, which is reassuring given that they are both trying to address the question of whether the data shows support for the characteristic coherent signature of a supermassive black hole binary. However, there are several advantages to the coherence test over the null distribution method: i) it has a significantly smaller computational cost; ii) it can quantify levels of preference \emph{against} the binary model (see Figure \ref{fig:bf_vs_pvalue}); iii) it is better defined, unlike the null distribution, which can change depending on the method used to generate it (see Figure \ref{fig:background_cw_fast_evolve_15psr}). As a result, we find the coherence test to be a more compelling option in practice.

In this study we have taken the first steps towards a robust detection of individual supermassive black hole binaries. We expect that in the coming years these methods will be improved upon and additional approaches will be developed as we prepare for the first detection. In particular, all our results are in the presence of white noise only for simplicity and computational efficiency. It will be important to understand how these tests work in the presence of red noise, and in the presence of a cross-correlated stochastic background. It will also be interesting to investigate how our ability to confidently detect a binary might be improved by more precise pulsar distance measurements. This is motivated by the fact that the model flexibility due to the unknown pulsar term phases is the main limitation of how confidently we can distinguish a coherent and an incoherent signal. Finally, it might be important to investigate how multiple binaries with similar frequencies and amplitudes might complicate the issue as they could blend together in a way that ruins the specific coherence one expects when modeling them as a single binary. 

\ack
We thank Rutger van Haasteren and Rand Burnette for feedback on the manuscript. 
Some of the results in this paper have been derived using the \texttt{healpy} \cite{healpy} and \texttt{HEALPix} \cite{healpix} package. We appreciate the support of the NSF Physics Frontiers Center Award PFC-2020265. S.R.T acknowledges support from NSF AST-2307719 and an NSF CAREER \#2146016.

\appendix
\section{Implied prior on per-pulsar amplitude in coherent model}
\label{sec:implied_prior}
The full coherent signal model from Section \ref{ssec:signal_model} describes the signal in each pulsar via a global model. However, in the low-frequency low-chirp-mass limit, where the Earth and pulsar terms have approximately the same frequency, the resulting signal in a given pulsar is simply a sine-wave (see Eq.~(\ref{eq:incoh_waveform})), which can be described by a frequency ($f_0$), an amplitude ($A_\alpha$), and an initial phase ($\Phi'_{\alpha}$). The frequency will be identical to the signal model frequency, and the prior distribution of the initial phase will be uniform between 0 and $2\pi$, if usual uninformative priors are chosen on all signal parameters. However, the resulting prior on $A_\alpha$ is nontrivial, and will be derived below.

The total signal in pulsar $\alpha$ can be expressed as \cite{nanograv_15yr_cw}:
\begin{equation}
    s_\alpha = F^+ \Delta s_+ + F^\times \Delta s_\times,
\end{equation}
where $\Delta s_{+,\times}=s_{+,\times}^{\rm PSR} - s_{+,\times}^{\rm Earth}$ is the signal difference between the pulsar and Earth terms for plus and cross polarizations, and $F^{+,\times}$ are the corresponding antenna pattern functions, which can be written without loss of generality for a pulsar towards the northern celestial pole as \cite{JustinCWMethods}:
\begin{eqnarray}
    F^+ &= \frac{1}{2} \cos 2\psi (1+\cos \theta) \, , \\
    F^\times &= \frac{1}{2} \sin 2\psi (1+\cos \theta) \, .
\end{eqnarray}

The Earth and pulsar terms signals are the same except for a $\Phi_{\alpha}$ phase offset, and they can be written as:
\begin{eqnarray}
    s_+ &= -\frac{A_{\rm E}}{2\pi f_{\rm E}} (1+ \cos^2 \iota)  \sin 2 \Phi(t)  \, , \\
    s_\times &= 2 \frac{A_{\rm E}}{2\pi f_{\rm E}} \cos \iota \cos 2 \Phi(t) \, .
\end{eqnarray}

If we expand the pulsar terms with trigonometric identities, we end up with three $\cos 2 \Phi(t)$ and three $\sin 2 \Phi(t)$ terms. Thus the total signal is a sine wave with an amplitude that can be expressed as the root sum square of the sine and cosine term amplitudes:
\begin{eqnarray}
    A_\alpha = \frac{A_{\rm E}}{2\pi f_{\rm E}} \frac{1+\cos \theta}{2} \sqrt{\lambda^2 + \kappa^2},
\end{eqnarray}
where:
\begin{eqnarray}
    \lambda &=  (1+\cos^2 \iota) \cos 2 \psi (1-\cos \Phi_\alpha) + 2 \cos \iota \sin 2 \psi \sin \Phi_\alpha  \, , \\
    \kappa &= (1+\cos^2 \iota) \cos 2 \psi \sin \Phi_\alpha - 2 \cos \iota \sin 2 \psi (1-\cos \Phi_\alpha) \, .
\end{eqnarray}

\begin{figure}[htbp]
\centering
\includegraphics[width=0.8\columnwidth]{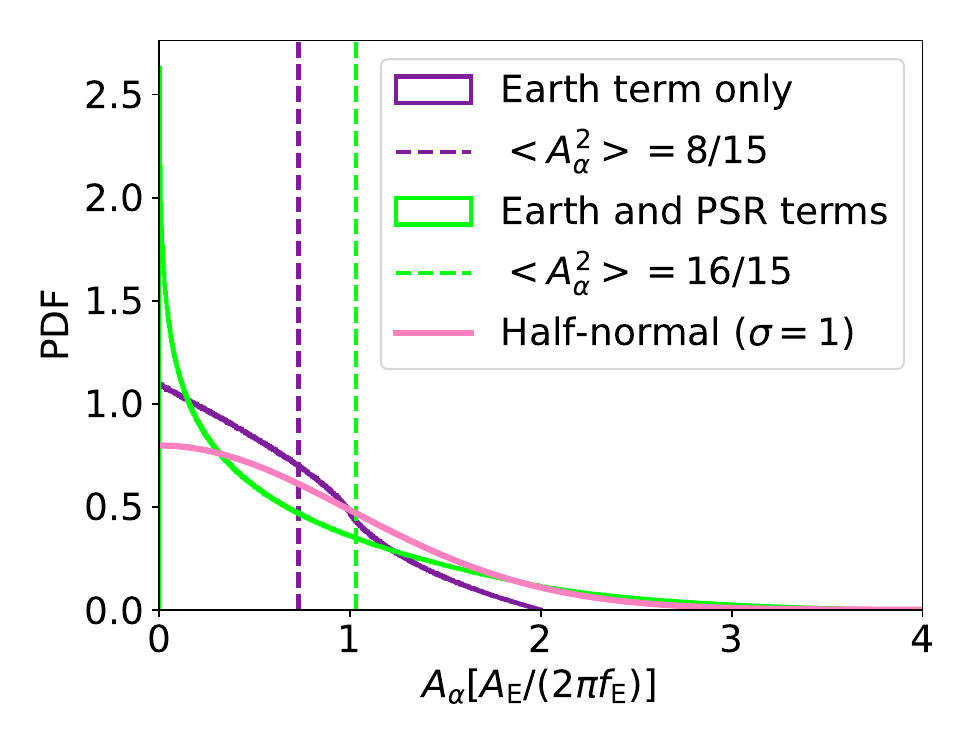}
\caption{Implied prior on per-pulsar amplitude under the coherent signal model with (green) and without (purple) pulsar terms. Dashed vertical lines indicate the average amplitudes, which agree well with analytical expressions for those. The half-normal prior used in the incoherent model is also show for comparison (pink).}
\label{fig:implied_prior}
\end{figure}

Thus for given $A_{\rm E}$ and $f_{\rm E}$, $A_\alpha$ is a random variable, which is a nonlinear function of four uniformly distributed random variables: $\cos \theta \in [-1,1]$, $\cos \iota \in [-1,1]$, $\psi \in [0,\pi]$, and $\Phi_\alpha \in [0,2\pi]$. The resulting distribution cannot be trivially expressed in closed form, but we can easily draw samples from it. Figure \ref{fig:implied_prior} shows this distribution both with (green curve) and without the pulsar terms (purple curve). We confirmed that the numerically calculated means from these agree with the analytic results for angle-averaged amplitudes (see e.g.~Ref.~\cite{hasasia}), which are shown as dashed vertical lines. The half-normal distribution used as a prior in our incoherent model (see Section \ref{ssec:coherence_test}) is also shown for comparison (pink curve). We can see that this has the same general appearance as the true implied distribution, as it peaks at zero and drops off with a long tail towards higher values. This validates this prior choice, but also indicates that it may be beneficial to investigate using the full implied prior in the future.

\section*{References}
\bibliographystyle{unsrt_et_al_15-3author_notitle}
\bibliography{cw_p_value_refs}

\end{document}